\documentstyle[seceq,epsf,wrapft,twoside]{ptptex}

\notypesetlogo  
\title{Candidates for Chiral Particles}
\author{%
Kenji {\sc Yamada}
}
\inst{%
Department of Engineering Science, Junior College Funabashi Campus,\\
Nihon University, Funabashi 274-8501, Japan
}

\abst{%
We find possible candidates for the new type of $ q\overline{q} $ meson, mesonic chiralons, 
whose existence is predicted in the covariant level-classification scheme of 
hadrons with $\widetilde{U}(12)_{SF}\otimes O(3, 1)_{L}$ symmetry, by assigning 
all the known light-unflavored and strange mesons, with masses below about 
1.8 GeV, to $\widetilde{U}(12)_{SF}\otimes O(3, 1)_{L}$ multiplets.
}

\begin{document}
\maketitle

\section{Introduction}

The covariant level-classification scheme of hadrons with 
$\widetilde{U}(12)_{SF}\otimes O(3, 1)_{L}$ symmetry has been proposed,\cite{ref:1} which 
gives a covariant quark representation for composite hadrons with definite 
Lorentz and chiral transformation properties. This covariant classification scheme unifies two 
seemingly contradictory viewpoints, nonrelativistic and extremely relativistic ones, 
of quark-model classification for hadrons. The one is based on 
the nonrelativistic quark model with ${SU}(6)_{SF}\otimes O(3)_{L}$ symmetry 
and the other stems from the fact that the lightest pseudoscalar mesons of 
the flavor octet ($\pi $, $K$, $\overline{K}$, $\eta$) can be identified with the 
Nambu-Goldstone bosons associated with spontaneous breaking of chiral 
symmetry, which plays an important role in the low-energy hadron physics.
In this scheme it is predicted that there exist additional  states of hadrons, 
called \textit{chiralons},
which do not appear in the conventional scheme based on nonrelativistic quark models.
In the case of mesons, there appear mesonic chiralons for light-quark $q\overline{q}$ 
and heavy-light-quark $Q\overline{q}$ and $q\overline{Q}$ systems,\footnote
{For experimental indications of the heavy-light-quark mesonic chiralons, 
see Ref. \citen{ref:10}.} 
while for heavy-quark $Q\overline{Q}$ systems the covariant level-classification scheme 
gives the same results on existing states as the conventional nonrelativistic 
classification scheme, so that there exist no mesonic chiralons.

Here we restrict our discussion to light-quark meson systems and in order to find 
possible candidates for mesonic chiralons we try to assign all the known light-unflavored 
and strange mesons with masses below about 1.8 GeV to the flavor $q\overline{q}$ nonets 
which exist in the covariant level-classification scheme.

\section{Listing of observed mesons below about 1.8 GeV}

We make a listing of all the observed light-unflavored and strange mesons with 
masses below about 1.8 GeV, which are expected to be composed of 
light-quarks ($u$, $d$, $s$), according to the Particle Data Group (PDG), the 2002 
edition of the \textit{Review of Particle Physics},\cite{ref:2} except for the following mesons:

\begin{itemize}
\item $\eta(1410)$ and $\eta(1475)$ \\ 
\hspace*{12pt} Since there is evidence for the existence of two psedoscalar mesons
 in the $\eta(1440)$ 
mass region,\footnote{See the ``Note on the $\eta(1440)$'' in the $\eta(1440)$ 
Particle Listings in Ref. \citen{ref:2}.} we consider that the PDG entry $\eta(1440)$ 
represents at least two states, which we call $\eta(1410)$ and $\eta(1475)$. 
The former decays into $K\overline{K}\pi$ mainly through $a_{0}(980)\pi$ (or directly), 
while the latter mainly through $K^{*}(892)\overline{K}$.
\end{itemize}

\begin{itemize}
\item $\kappa(900)$ \\ 
\hspace*{12pt} In recent years there has been reported evidence for the existence of 
the light scalar meson, $\kappa(900)$, in the reanalyses\cite{ref:3} 
of $K\pi$ scattering phase-shift data, 
the Dalitz-plot analysis\cite{ref:4} of the decay $D^{+} \to K^{-}\pi ^{+}\pi^{+}$ 
by the Fermilab E791 Collaboration, and the analysis\cite{ref:5} of the decay 
$ J/\psi \to K^{*}(892)K\pi $ by the BES Collaboration. We consider the 
$\kappa(900)$ to be a genuine resonance.
\end{itemize}

\begin{itemize}
\item $\rho (1250)$ and $\omega (1200)$ \\ 
\hspace*{12pt} There have been reported\cite{ref:6} several experimental indications of the existence 
of a low-mass isovector vector meson, the $\rho (1250)$, at a mass of 
around 1200 to 1300 MeV.\footnote{See also the $\rho(1450)$ Particle Listings and 
the ``Note on the $\rho(1700)$'' in the $\rho(1700)$ Particle Listings in Ref. \citen{ref:2}.}
For a low-mass isoscalar vector meson the existence of the $\omega(1200)$ 
has been claimed\cite{ref:7} in the analysis of 
the $e^{+}e^{-} \to \pi ^{+}\pi ^{-}\pi ^{0}$ cross section by the SND Collaboration. 
We now assume the existence of these two vector-meson states to be true.
\end{itemize}
The resulting meson listing is shown in Table I. Here we add some comments 
on this meson listing:
\begin{enumerate}
\renewcommand{\labelenumi}{(\theenumi)}
\item In the isoscalar sector there is at least an extra psedoscalar meson, 
one of the three states $\eta(1295)$, $\eta(1410)$, and $\eta(1475)$, 
which has no place in $q\overline{q}$ nonets within the nonrelativistic classification scheme.
\item We have several scalar mesons, such as the $\sigma(600)$, $f_{0}(980)$, 
$a_{0}(980)$, and $\kappa(900)$, which are difficult to be interpreted as 
nonrelativistic $q\overline{q}$ states.
\item There are some vector mesons with the normal $J^{PC}$ 
quantum numbers, which are unnatural to assign to $q\overline{q}$ states 
within the nonrelativistic classification scheme.\footnote{For a recent review, 
see, for example, Ref. \citen{ref:8}.}
\item There are two exotic mesons, the $\pi _{1}(1400)$ and $\pi _{1}(1600)$,  
with unusual quantum numbers $I^{G }(J^{PC}) = 1^{-}(1^{-+})$,
 which are not accessible to nonrelativistic $q\overline{q}$ systems.
\item Since an assignment to the conventional $1^{3}P_{2}$ $q\overline{q}$ meson nonet 
with $J^{PC}= 2^{++}$ is completed unambiguously by the tensor mesons 
$f_{2}(1270)$, $f_{2}'(1525)$, $a_{2}(1320)$, and $K_{2}^{*}(1430)$, 
any other $2^{++}$ mesons below about 1.6 GeV, 
such as $f_{2}(1430)$ and $f_{2}(1565)$, seem unable to be 
nonrelativistic $ q\overline{q} $ states.
\end{enumerate}

\section{Possible assignments for the listed mesons in the covariant level-classification scheme}

In the covariant level-classification scheme of hadrons we have
additional $q\overline{q}$ meson nonets, which do not appear in the nonrelativistic classification scheme, with normal as well as exotic $J^{PC}$ quantum numbers. 
The predicted $J^{PC}$ quantum numbers of $q\overline{q}$ nonets 
with $L= 0, 1, 2$ ($L$ being the relative orbital angular momentum between 
a quark and an antiquark) are as follows: 
\begin{itemize}
\item $L = 0$ nonets
	\begin{itemize}
	\renewcommand{\labelitemii}{}
	\item $J^{PC} = 0^{-+} (2), 1^{--} (2), 0^{++}, \underline{0^{+-}}, 1^{++}, 1^{+-}$
	\end{itemize}

\item $L = 1$ nonets
	\begin{itemize}
	\renewcommand{\labelitemii}{}
	\item $J^{PC} = 1^{+-} (2), 0^{++} (2), 1^{++} (2), 2^{++} (2), \underline{0^{--}},
	 0^{-+}, 1^{--} (2), \underline {1^{-+}} (2), 2^{--}, 2^{-+}$
\end{itemize}

\item $L = 2$ nonets
	\begin{itemize}
	\renewcommand{\labelitemii}{}
	\item $J^{PC} = 2^{-+} (2), 1^{--} (2), 2^{--} (2), 3^{--} (2), 1^{++}, 1^{+-}, 2^{++} (2), 
	\underline {2^{+-}} (2), 3^{++}, 3^{+-}$
	\end{itemize}
\end{itemize}
Here the numbers, unless one, in parentheses are those of predicted nonets 
with respective $J^{PC}$
and the exotic quantum numbers are underlined. 
We have the two pseudoscalars $P(0^{-+})$ and $\widetilde{P}(0^{-+})$, two scalars 
$S(0^{++})$ and $\widetilde{S}(0^{+-})$, two vectors $V(1^{--})$ 
and $\widetilde{V}(1^{--})$, and two axial-vectors 
$A(1^{++})$ and $\widetilde{B}(1^{+-})$ for the ground-state ($L= 0$) nonets. 
In the nonrelativistic limit of an effective quark mass becoming large,  
where the chiral symmetry is lost,
the additional six states $\widetilde{P}$, 
$S$, $\widetilde{S}$, $\widetilde{V}$, 
$A$, and $\widetilde{B}$ disappear and 
the pseudoscalar and vector states, $P$ and $V$, still survive, 
in accord with the nonrelativistic level-classification scheme.
It is noted that the chiral symmetry might not be effective any more
 for the excited states, especially orbital $L \geq 2$ and radial $n_r \geq 1$ excitations, 
of the additional six ground-state nonets.\footnote{For an inference about 
the validity of the effective chiral symmetry, see Ref. \citen{ref:9}.}  
In this case such excited-state nonets do not appear and 
there exist only the excitations of the conventional ground-state 
$P$ and $V$ nonets for $L \geq 2$ and $n_r \geq 1$.

 We now assign tentatively all the listed mesons in Table I to 
 the above $ q\overline{q} $ nonets and radial excitations of 
 the $L = 0$ nonets, resorting to their quantum numbers and masses. 
The resulting assignments are shown in Table II. From this table 
we see that there is enough room to be assigned, though 
their assignments are ambiguous, and there are quite
suitable places, in paticular, for the light scalar nonet $[\sigma(600)$, $f_{0}(980)$, 
$a_{0}(980)$, $\kappa(900)]$ as well as the two exotic 1$^{-+}$ mesons 
$\pi _{1}$(1400) and $\pi _{1}$(1600).

\section{Concluding remarks}

We have presented the possible assignments for all the observed mesons 
below about 1.8 GeV in the covariant level-classification scheme of hadrons. 
To establish their assignments we need to examine further mass spectra and decay properties 
of the predicted states. 
It is also interesting to search for exotic mesons with 
$J^{PC} = 0^{+-}, 0^{--}, 1^{-+}, 2^{+-}$ below about 1.8 GeV
 and new tensor $2^{++}$ mesons below about 1.6 GeV
 other than the established $1^{3}P_{2}$ $q\overline{q}$ meson nonet.


\newpage

\begin{figure}[hbpt]
	\begin{description}	
	\item Table I. \ \ The particle listing of mesons with masses below about 1.8 GeV. 
	The data are taken from Ref. \citen{ref:2}, unless otherwise noted in the text. 
	\end{description}
	\vspace{3pt}
	\begin{center}
	\epsfxsize =13.5cm
	\centerline{\epsfbox{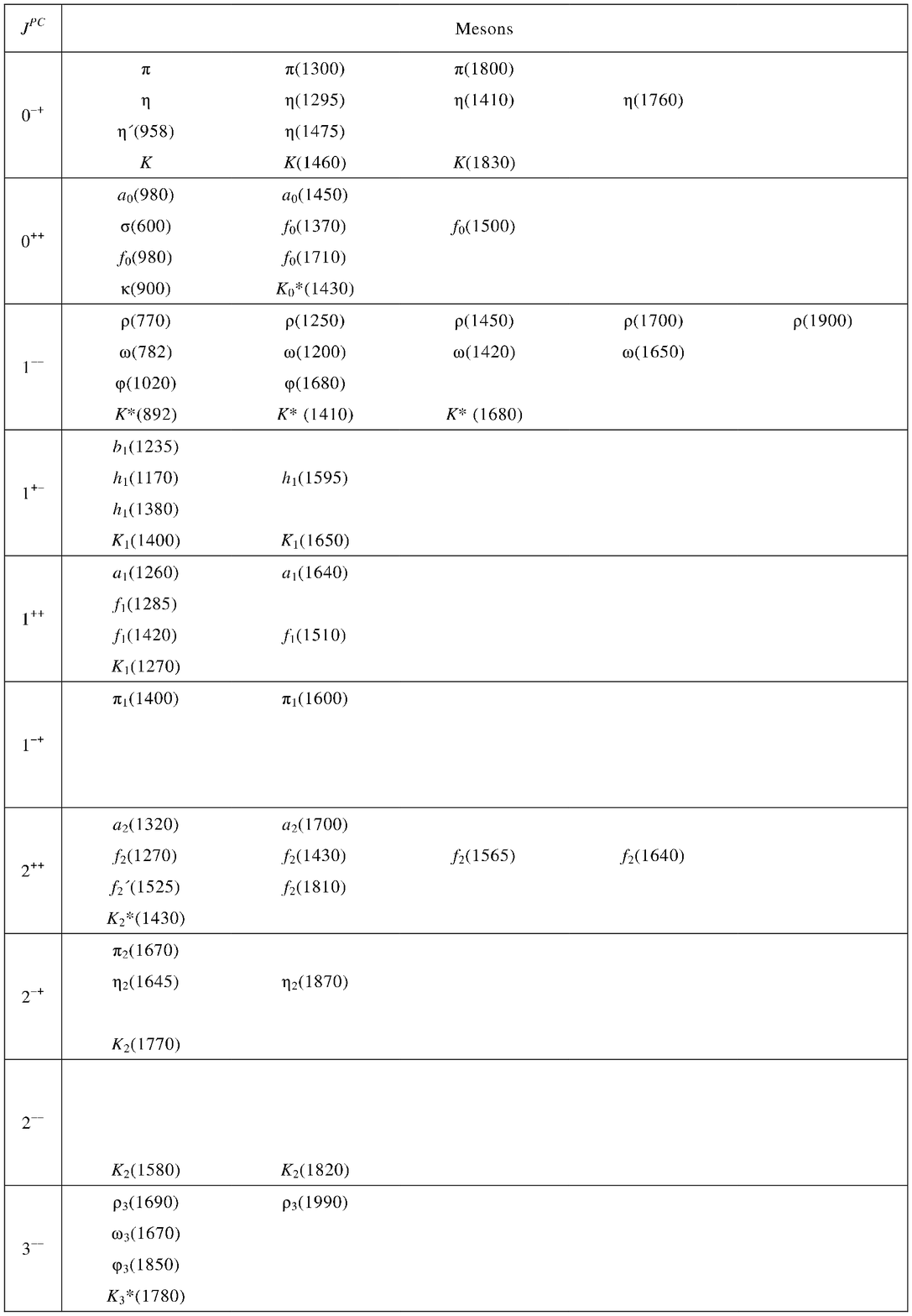}}
	\end{center}
\end{figure}%

\newpage

\begin{figure}[hbpt]
	\begin{description}	
	\item Table II. \ \ Possible assignments for all of the known mesons with masses below 
	about 1.8 GeV in the covariant level-classification scheme. 
	\end{description}
\vspace{-40pt}
	\begin{center}
	\epsfxsize =17cm \centerline
	{\epsfbox{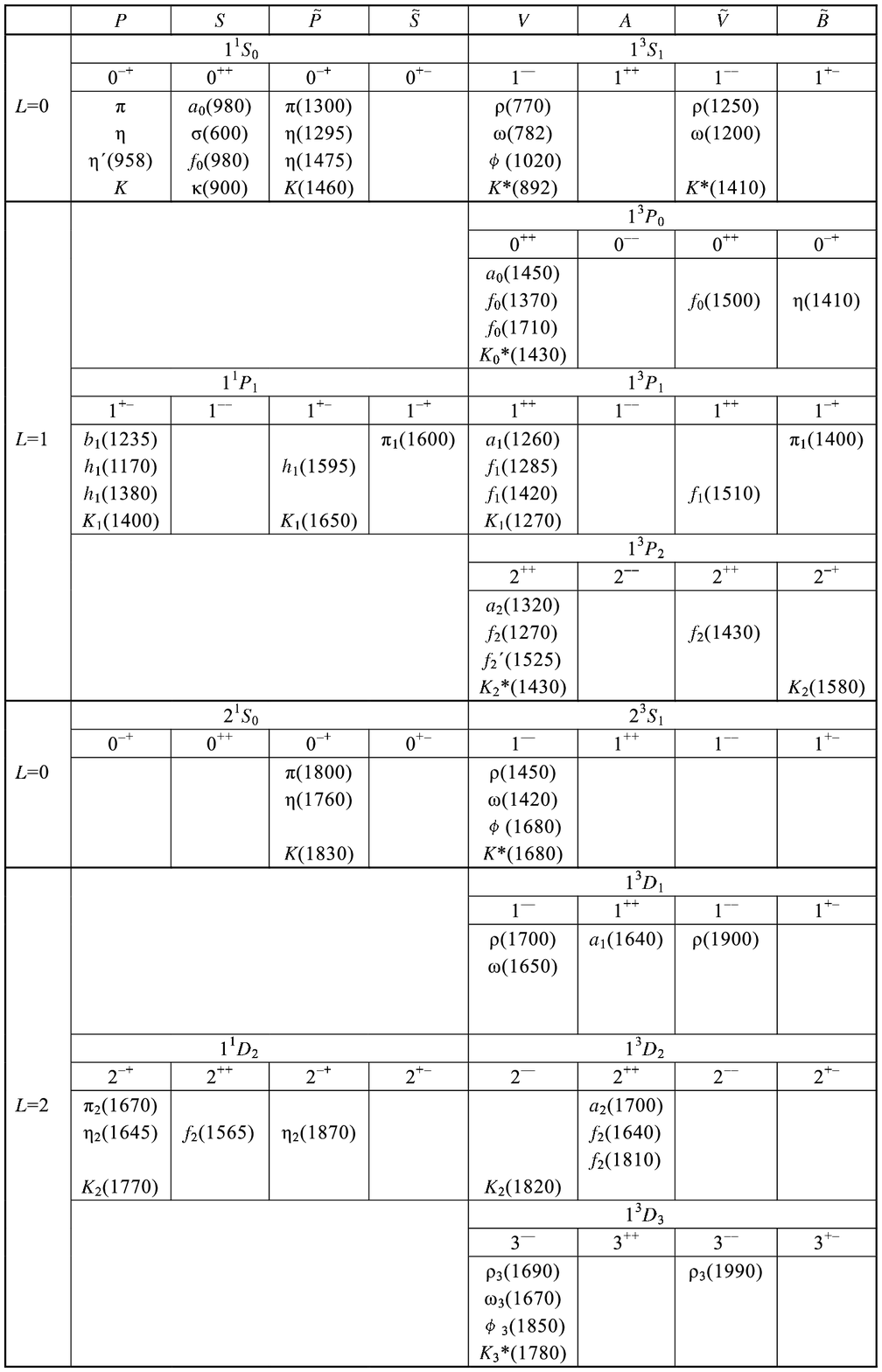}}
	\end{center}
\end{figure}


\begin{thebibliography}{99}
\bibitem{ref:1}
	S. Ishida, M. Ishida and T. Maeda, Prog. Theor. Phys. \textbf{104} (2000), 785.\\
	S. Ishida and M. Ishida, Phys. Lett. B \textbf{539} (2002), 249.\\
	S. Ishida, these proceedings.
\bibitem{ref:10}
	K. Yamada et al., in \textit{HADRON SPECTROSCOPY}, Proc. of the Ninth
	International Conference on Hadron Spectroscopy, IHEP, Protvino, Russia, 2001, 
	ed. D. Amelin and A. M. Zaitsev, AIP Conf. Proc., Vol. 619 (AIP, New York, 2002),
	p. 657.\\
	D. Ito et al., \PTP{108,2002,953}.\\
	I. Yamauchi, these proceedings.
\bibitem{ref:2}
	Particle Data Group, K. Hagiwara et al., Phys. Rev. D \textbf{66} (2002), 010001.
\bibitem{ref:3}
	E. Beveren et al., Z. Phys. C \textbf{30} (1986), 615. \\
	S. Ishida et al., Prog. Theor. Phys. \textbf{98} (1997), 621. \\
	D. Black et al., Phys. Rev. D \textbf{58} (1998), 054012.
\bibitem{ref:4}
	E791 Collaboration, E. M. Aitala et al., Phys. Rev. Lett. \textbf{89} (2002), 121801. \\
	C. G\"{o}bel, these proceedings.
\bibitem{ref:5}
	N. Wu, these proceedings. \\
	T. Komada, these proceedings.
\bibitem{ref:6}
	D. Aston et al., SLAC-PUB-5606 (1994).\\
	OBELIX Collaboration, A. Bertin et al., Phys. Lett. B \textbf{414} (1997), 220.\\
	P. Frenkiel et al., Nucl. Phys. B \textbf{47} (1972), 61.\\
	J. Ballam et al., Nucl. Phys. B \textbf{76} (1974), 375.
\bibitem{ref:7}
	 M. N. Achasov et al., Phys. Lett. B \textbf{462} (1999), 365.
\bibitem{ref:8}
	A. Donnachie and Yu. S. Kalashnikova, in \textit{HADRON SPECTROSCOPY}, Proc. of the Ninth
	International Conference on Hadron Spectroscopy, IHEP, Protvino, Russia, 2001, 
	ed. D. Amelin and A. M. Zaitsev, AIP Conf. Proc., Vol. 619 (AIP, New York, 2002),
	p. 5.
\bibitem{ref:9}
	S. Ishida and M. Ishida, in \textit{HADRON SPECTROSCOPY}, Proc. of the Ninth
	International Conference on Hadron Spectroscopy, IHEP, Protvino, Russia, 2001, 
	ed. D. Amelin and A. M. Zaitsev, AIP Conf. Proc., Vol. 619 (AIP, New York, 2002),
	p. 187.

\end{thebibliography}
\end{document}